\documentclass{aa}  

\usepackage{graphicx}
\usepackage{txfonts}
\usepackage{comment}
\usepackage{amsmath}
\usepackage{float}

\usepackage{lineno}
\begin{document} 

   \title{Image calibration between the Extreme Ultraviolet Imagers on Solar Orbiter and the Solar Dynamics Observatory}
   \titlerunning{Instrument intercalibration between Solar Orbiter and the Solar Dynamics Observatory}

   \author{C. Schirninger
          \inst{1}
          \and
          R. Jarolim\inst{2}
          \and
          A. M. Veronig\inst{1,}\inst{3}
          \and
          A. Jungbluth\inst{4}
          \and
          L. Freischem\inst{5}
          \and
          J.E. Johnson\inst{6}
          \and
          V. Delouille\inst{7}
          \and
          L. Dolla\inst{7}
          \and
          A. Spalding\inst{8}
          }

   \institute{Institute of Physics, University of Graz,
              Universitätsplatz 5, 8010 Graz, Austria
         \and
             High Altitude Observatory, National Center for Atmospheric Research, 3080 Center Green Dr, Boulder, USA  
         \and
             Kanzelhöhe Observatory for Solar and Environmental Research, University of Graz, 
             Treffen am Ossiacher See, Austria
        \and
            European Space Agency (ESA)-ECSAT, Fermi Avenue, Harwell, UK
        \and
            Department of Physics, University of Oxford, Oxford, UK
        \and 
            Image Processing Laboratory (IPL) Department Universitat de València, València, Spain
        \and
            Solar-Terrestrial Centre of Excellence (SIDC), Royal Observatory of Belgium, Ringlaan 3 Av. Circulaire, 1180 Brussels, Belgium
        \and
            Trillium Technologies Inc. (Trillium USA), 8668 John Hickman Pkwy, STE 301 Frisco, TX
            }

 
  \abstract
   {To study and monitor the Sun and its atmosphere, various space missions have been launched in the past decades. With the rapid improvement in technology and different mission requirements, the data products are subject to constant change. However, for long-term studies such as solar variability or multi-instrument investigations, uniform data series are required.
   
   In this study, we build on and expand the Instrument-to-Instrument translation (ITI) framework, which provides unpaired image translations. We apply the tool to data from the Extreme Ultraviolet Imager (EUI), specifically the Full Sun Imager (FSI) on  Solar Orbiter (SolO) and the Atmospheric Imaging Assembly (AIA) on the Solar Dynamics Observatory (SDO). This approach allows us to create a homogeneous data set that combines the two extreme ultraviolet (EUV) imagers in the 174/171~{\AA} and 304~{\AA} channels.
   
   We demonstrate that ITI is able to provide image calibration between SolO and SDO EUV imagers, independent of the varying orbital position of SolO. The comparison of the intercalibrated light curves derived from 174/171~{\AA} and 304~{\AA} filtergrams from EUI and AIA shows that ITI can provide uniform data series that outperform a standard baseline calibration. We evaluate the perceptual similarity in terms of the Fr\'{e}chet Inception Distance (FID), which demonstrates that ITI achieves a significant improvement of perceptual similarity between EUI and AIA. The study provides intercalibrated observations from SolO/EUI/FSI with SDO/AIA, enabling a homogeneous data set suitable for solar cycle studies and multi viewpoint investigations.
}


   \maketitle
%

\section{Introduction}
The Solar Orbiter (SolO; \citealp{Mueller2020}) mission, launched in 2020, has already provided exceptional insight in the study of solar flares, coronal mass ejections (CMEs), coronal holes and very small brightenings with imagers in the extreme ultraviolet (EUV) and X-ray (e.g., \citealp{Collier2024, Chitta2023, Berghmans2021, Schwanitz2023, Kane2021, purkhart2024, podladchikova2024, saqri2024}). In particular, the Extreme Ultraviolet Imager (EUI; \citealp{Rochus2020}) observes of the solar atmosphere in the EUV, globally as well as at high spatial- and temporal resolution. With its highly elliptical orbit and the out-of-ecliptic transition in 2025, SolO provides the perfect conditions to analyze eruptive phenomena from different vantage points. On the other hand, the Solar Dynamics Observatory (SDO; \citealp{Pesnell2012}) launched in a geosynchronous orbit around Earth in 2010 provides a long database of solar observations. In particular with its high temporal and spatial resolution imaging capabilities from the Atmospheric Imaging Assembly (AIA; \citealp{Lemen2011}), SDO provides detailed insights into dynamic phenomena such as solar flares. 

Long term solar monitoring is important to understand solar activity and consequently solar eruptive events as these pose a significant threat not only to Earth and its environment but also to our assets in space. The sunspot number forms the longest directly observed index for solar cycle and variability studies \citep{Usoskin2003}. On the other hand, the solar EUV radiation is driving the variability in Earth's upper atmosphere, affecting satellites, communication and navigation \citep{Woods2012}. As more space missions are launched, we are improving our understanding of the Sun's atmosphere and therefore its eruptive phenomena. However, with the rapid improvement in technology and also different mission requirements, the instruments and therefore the data products are changing. The calibration of these instruments ensures a direct intensity comparison, reducing the potential for bias caused by differences in instrumentation and also changes due to degradations in space. For multi-instrument, solar cycle studies and long term solar monitoring uniform data series are required. Furthermore, the lifetime of these space missions is limited. Consequently, image calibration bridge the gap between older and newer missions to build a uniform data series.

Traditional image calibration methods for EUV observations are based on standardization by subtracting the average logarithmic intensity and dividing by the standard deviation of the logarithmic intensity of these observations, as shown by \cite{Hamada2020}. \cite{DosSantos2021} first applied deep neural networks using convolutional neural networks (CNNs) for autocalibration including instrument degradation correction of the AIA instrument. Previous studies have also used deep neural networks for image enhancement \citep{Baso2018} and image denoising \citep{Baso2019}. \cite{youn2025} applied Generative Adversarial Networks (GANs) to synthesize the remaining AIA channels on SolO in order to perform a differential emission measure (DEM) analysis.

In order to obtain joint data products of the EUV imagers onboard SolO and SDO, we applied the Instrument-to-Instrument (ITI) translation tool \citep{Jarolim2025}. ITI has been shown to be a universally applicable tool for solar image data, allowing image enhancement, image calibration, and super-resolution observations. In addition, ITI has been applied to ground-based solar data to mitigate seeing effects from the Kanzelhoehe Solar Observatory (KSO; \citealp{Potzi2021}) and to reconstruct high-resolution ground-based solar observations from the GREGOR telescope \citep{Schirninger2024}.
Our aim is to achieve simultaneous image calibration and image enhancement by applying the ITI tool to translate EUV observations from the EUI instrument on SolO into corresponding observations from SDO/AIA. 

An important property of ITI is that it utilizes unpaired image-to-image translation, which does not require a spatial nor temporal overlap between the two data sets. This is particularly important for the calibration between SolO and SDO, due to the sparse set of aligned observation. In addition, differences in instrument resolution, wavelength range, filter characteristics and quality are complex and a calibration based on statistical quantities is typically insufficient \citep{Hamada2020}. The neural network is learning the image distribution of the ground-truth reference data set and therefore allows to translate between two instruments. This provides joint data products, to be used for detection methods e.g. coronal hole detections \citep{Caplan2016, Reiss2021, Jarolim2021}, flare detections \citep{Nistico2017}, triangulation methods as shown by \cite{Aschwanden2008_triang}, or to obtain a 3D representation of the Sun from multi-viewpoint observations using neural radiance fields (SuNeRF; \citealp{Jarolim2024}), where ITI was also already used.

In Sect.\,\ref{sec:data} we introduce the data used in this study as well as the pre-processing steps to achieve machine learning ready data sets. Section\,\ref{sec:method} provides an overview of the method. The results of the study are discussed in Sect.\,\ref{sec:results} including a quantitative comparison in Sect.\,\ref{subsec:quantcomp}, a qualitative comparison in Sect.\,\ref{subsec:qualcomp}, a comparison of translations for different distances of SolO to the Sun in Sect.\,\ref{subsec:distance} and a multi-viewpoint application in Sect.\,\ref{subsec:multiview}. In Sect. \ref{sec:conclusion} we present our conclusions from this study.

\section{Data}
\label{sec:data}

In this study, we apply the ITI tool on EUV observations from the SolO/EUI instrument and use observations from SDO/AIA as the calibration target for image calibration. The EUI instrument is equipped with two EUV imagers, the Full Sun Imager (FSI) taking full Sun observations and the High resolution Imager (HRIEUV) providing high resolution observations of a smaller field of view (FOV) of $(1~R_{\odot})^2$ at 1 AU and $(0.28~R_{\odot})^2$ perihelion, respectively \citep{Rochus2020}. We perform the calibration between the 174~{\AA} and 304~{\AA} filters from SolO/EUI/FSI and the 171~{\AA} and 304~{\AA} filters of SDO/AIA. Note that the AIA 171~{\AA} and the FSI 174~{\AA} channel observe emission from different iron ions, with AIA 171~{\AA} primarily sensitive to Fe~IX and FSI 174~{\AA} dominated by Fe~X. Consequently, differences between images from these two channels are expected due to their different temperature sensitivity.

\begin{table}[h!tbp]
\caption{Number of observation of the data set split into training, validation and three different test sets.}       
\label{table:dataset}  
\centering                         
\begin{tabular}{c | c c c}       
\hline\hline                               
& EUI/FSI & & AIA \\
\hline \\
Training set & 2306 & & 3428 \\
& & &  \\
Validation set & 205 & & 776 \\ 
& & &  \\
Test set 1 (Earth aligned) & 89 & & 89 \\
& & & \\
Test set 2 (Variable Distance) & 649 & & - \\                      
& & & \\
Test set 3 (Multi-viewpoint) & 81 & & 81  \\
& & & \\
Test set 4 (Lightcurve) & 3366 & & 3366 \\ 
\hline 
\end{tabular}
\end{table}

\subsection{FSI data}
\label{sec:fsi_data}

The FSI instrument provides full Sun observations in two wavelength channels, 174~{\AA} and 304~{\AA}. With a FOV of two solar diameters at perihelion, the full solar disk including the extended solar corona can be observed when the spacecraft is pointed at the solar limb. The angular resolution of the telescope is 10 arcsec. Since the orbit of SolO is very elliptical, the spatial resolution of the observations is changing according to the distance to the Sun. At perihelion, SolO covers a smaller region on the Sun and consequently achieves the highest resolution observations. We consider all observations available from the perihelion at $\sim$0.3 AU to the aphelion at $\sim$1.1 AU. Therefore, we aim to achieve an image calibration that also accounts for the varying resolution, due to the change in orbital position. The data used in this work are obtained from Data Release 6 \citep{euidatarelease6}.

To obtain a machine learning ready data set, we pre-process the FSI data by cropping all observations to a FOV of 1.1 solar radii and scale them to an image size of $1024 \times 1024$ pixels. Cropping the full Sun observations to 1.1 solar radii prevents the sampling of patches that are completely off-limb, since training is performed using image patches. Furthermore, the pixel values are mapped to an interval between $[-1, 1]$ using a normalization between the minimum intensity value and 90$\%$ of the maximum intensity value of the entire training data set. The FSI 174\,\AA~and 304\,\AA\,channel shows significant degradation. We apply a degradation correction by estimating the quiet Sun region over the entire training set from January 2022 to May 2023 and fit the degradation as a first-order polynomial (cf.~\citealp{Jarolim2025}). The fit parameters extracted from the linear fit are used to correct for the degradation of the entire data set including data after 2023. Consequently, the linear fit serves as a degradation approximation for later times. We use the linear fit to avoid filtering out solar cycle effects. The degradation and the linear fit to the data are shown in the  Appendix (Fig.\,\ref{fig:deg_corr}). The FSI data set spans from January 2022 to February 2023. We note that the SolO mission is conducting Solar Orbiter Observing Plans (SOOPs; \citealp{Zouganelis2020}). The SOOPs are designed to understand how the Sun generates and controls the heliosphere from a combined in-situ and remote sensing perspective. However, this also affects other instruments on the spacecraft. For some SOOPs, the imagers are not pointed at the center of the solar disk. This leads to intensity enhancements in the FSI data. Consequently, we remove all off-pointing SOOP observations from our data set. 

\begin{figure}[h!tbp]
   \centering
   \includegraphics[scale=0.058]{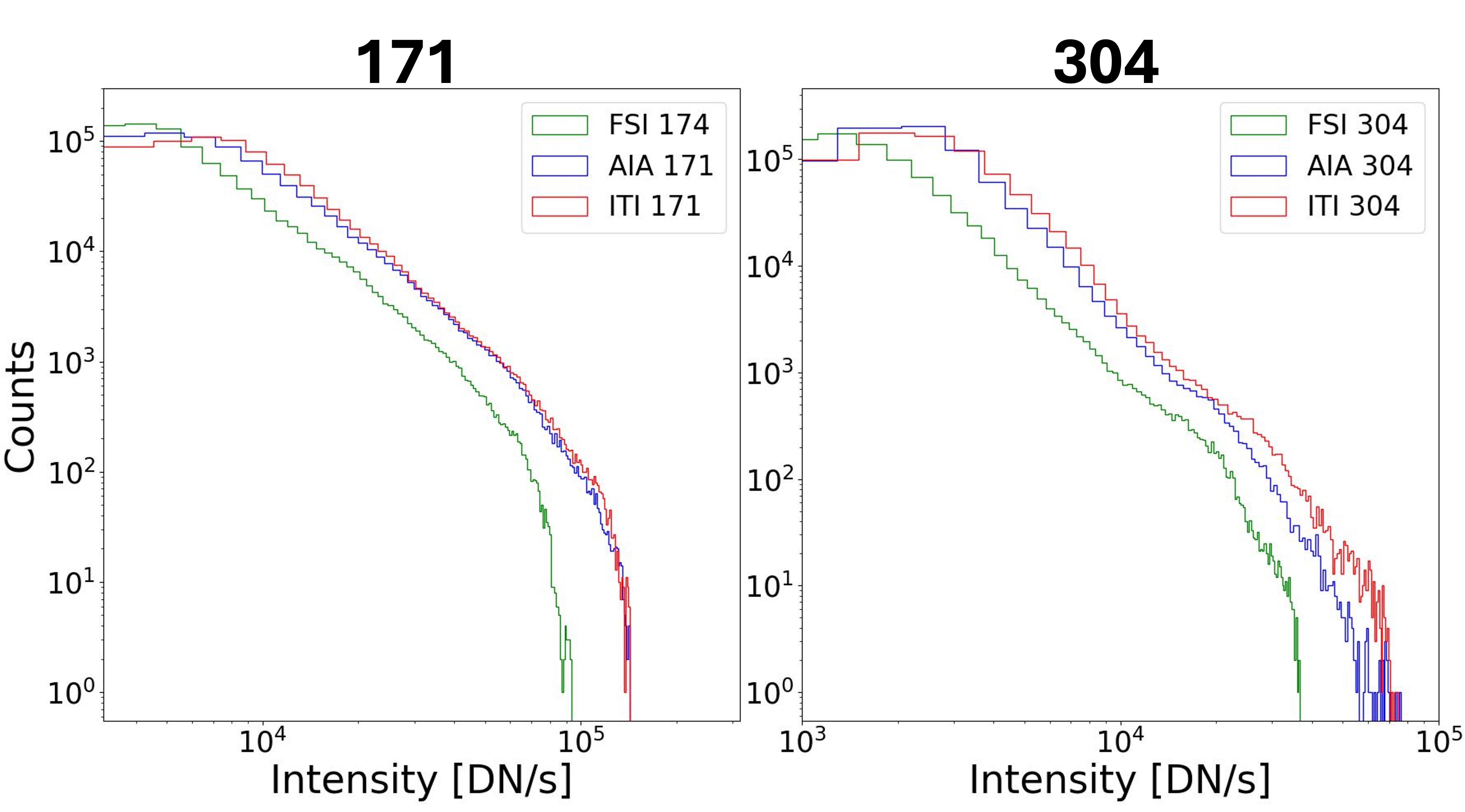}
   \caption{Comparison of the logarithmic intensity distributions over test set 1 of the baseline calibrated FSI observations (green), the ITI translations (red) generated from the FSI observations with AIA observations as target, and the reference AIA observations (blue) for 171~{\AA} (left) and 304~{\AA} (right), respectively.}
\label{fig:histogram}
\end{figure}

\subsection{AIA data}

For the reference data set, which is our calibration target, we utilize observations from SDO/AIA in the 171~{\AA} and 304~{\AA} filters. The SDO spacecraft operates in a geosynchronous orbit around Earth at approximately 1 AU. The AIA instrument provides high-resolution observations in the EUV, with a spatial resolution of 1.5 arcseconds and a temporal cadence of 12 seconds. AIA's field of view spans 1.3 solar radii, capturing both the full solar disk and the surrounding solar corona. Additionally, AIA offers a substantial database with observations starting from 2010. For our data set, we consider one observation per day, starting from May 2011 to April 2024. We perform a visual inspection of the data set and remove invalid observations (i.e., eclipse times, cosmic particles or instrument malfunctioned observations). Additionally, observations containing flares were excluded and are not considered for image calibration.
Analogous to the pre-processing of the FSI observations, the AIA data are pre-processed by cropping all observations to 1.1 solar radii and mapping the pixel values to an interval between $[-1, 1]$ using the minimum intensity value and 90$\%$ of the maximum intensity value of the entire training data set. Furthermore, the observations are downsampled from $4096\times 4096$ pixels to a resolution of $1024\times 1024$ pixels. Additionally, we correct for the time dependent degradation of the AIA instrument using the correction table according to \cite{Barnes2020}.

\begin{figure*}[h!tbp]
   \centering
   \includegraphics[scale=0.1]{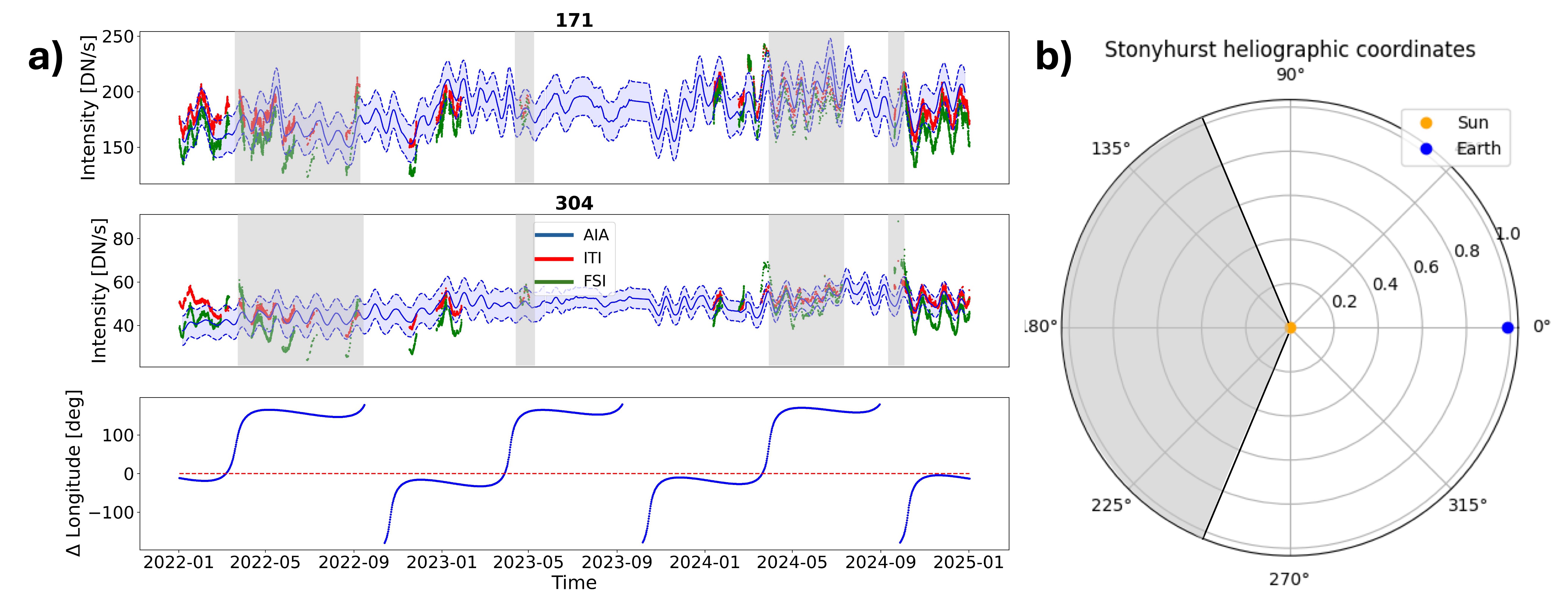}
   \caption{Comparison between the light curves from January 2022 to January 2025 derived from the baseline-calibrated FSI data (green), the ITI translated data (red) and the reference AIA data (blue), including training data. The blue shaded area corresponds to $\pm~1\,\sigma$ error range from AIA. In a) the first panel corresponds to the 171~{\AA} and the second to the 304~{\AA} channel. The light curves were co-aligned in longitude using the separation angle between the SDO and SolO spacecraft shown in the third panel. In b) the coordinate system is shown with the gray shaded area corresponding to the gray shaded area in a) indicating SolO was located on the solar far-side and thus not observing the same regions of the Sun as SDO.}
\label{fig:lightcurve}
\end{figure*}

\subsection{Data sets}
\label{sec:datasets}
In total, the FSI data set consists of 3699 observations in the time span from January 2022 to February 2023 for the 174~{\AA} and 304~{\AA} filters and 4607 observations from May 2011 to April 2024 for the AIA reference data set in both 171~{\AA} and 304~{\AA}, respectively (see Tab.\,\ref{table:dataset}). For training, we split the data set into training and validation sets, using data from February to October for training and observations from November and December for validation. This temporal separation reduces the risk of overlap between training and validation samples, which can occur due to the strong similarity of temporally adjacent observations \citep{Kim2019}. 

For our evaluation we prepare four test sets. The first test set consists of observations when SolO was close to the Sun-Earth line within 0.8 degrees longitude. It covers the period from 11 November 2021 to 14 November 2021 and 20 March 2024 from 13:00 to 17:00 UT with a cadence of one hour. In this time frame no off-pointing was performed for SolO and therefore allows for a direct comparison with the AIA observations that we use as "ground truth". There is a small latitude difference of $\sim$2.9$^\circ$ on average between SolO and SDO, making pixel-based metrics and intensity ratio images inappropriate for comparison. We refer to this data set as test set 1 (Earth aligned). The second test set is to compare how the distance of SolO to the Sun affects the ITI translations, containing observations form 0.3 AU to 0.9 AU. It covers the time range from 15 May 2023 to 28 May 2023 and February 2024 to March 2024. We refer to this data set as test set 2 (Variable Distance). The third test set consists of observations where SolO is not in the Sun-Earth line but has an overlap in the FOV with SDO/AIA to investigate whether ITI is capable of image calibration for multi-viewpoint studies. This test set covers the period 21 October 2023 to 24 October 2023. This test set allows to visually compare our calibrations with the ground truth AIA observations. We refer to this data set as test set 3 (Multi-viewpoint). The separation angle of the two spacecraft in longitude across test set 3 spans from $40^\circ$ to $33^\circ$, where SolO is West in longitude from SDO. Test set 4 (Lightcurve) is used to perform a statistical light curve analysis of our ITI translations and the baseline calibration. It covers the period from March 2023 to January 2025. All observations in the time frame of the test sets including ten days before and after were removed from the training and validation set.

\section{Method}
\label{sec:method}

We use the ITI tool to calibrate the EUI/FSI instrument onboard SolO with SDO/AIA. The tool was originally developed by \cite{Jarolim2025} and is provided as open source framework\footnote{https://github.com/spaceml-org/InstrumentToInstrument/}. Previously, ITI has been applied to data from SDO, the Solar Optical Telescope (SOT; \citealp{Tsuneta2008}) onboard Hinode, the Solar and Heliospheric Observatory (SOHO; \citealp{Domingo1995}) and the Solar Terrestrial Relations Observatory (STEREO; \citealp{Driesman2008}) to obtain image calibration, image enhancement and super-resolution observations.

The ITI tool is based on Generative Adversarial Networks (GANs; \citealp{Goodfellow2014}) and uses unpaired image-to-image translation \citep{Zhu2017}. The standard GAN setup consists of two competing networks, a generator network and a discriminator network. The aim of the generator network is to synthesize images in the reference domain so that the discriminator can no longer distinguish between real and synthesized images. Compared to paired image translation \citep{Isola2017}, no spatial or temporal overlap of the two data sets is required. The generator network learns the distribution of the reference data set by mapping images from a domain A to the reference domain B. The generator networks use a U-net architecture with downsampling and upsampling blocks and skip connections that help the network to use the information from the downsampling steps \citep{Ronneberger2015}. The discriminator networks consist of three individual networks, each of which performs downsampling steps. This allows optimization to be carried out at multiple resolution scales \citep{Wang2018}. 

In total, two discriminators and two generators are used for training. One training cycle involves translation from one instrument of domain A to the reference instrument of domain B and back to domain A (A-B-A). In a second cycle, the same translation is performed, but from domain B to domain A and back to B (B-A-B). The last two training cycles involve identity mappings using generator AB to translate images from domain B to domain B (B-B) and vice versa for images of domain A using generator BA (A-A).

\begin{figure*}[h!tbp]
   \centering
   \includegraphics[scale=0.12]{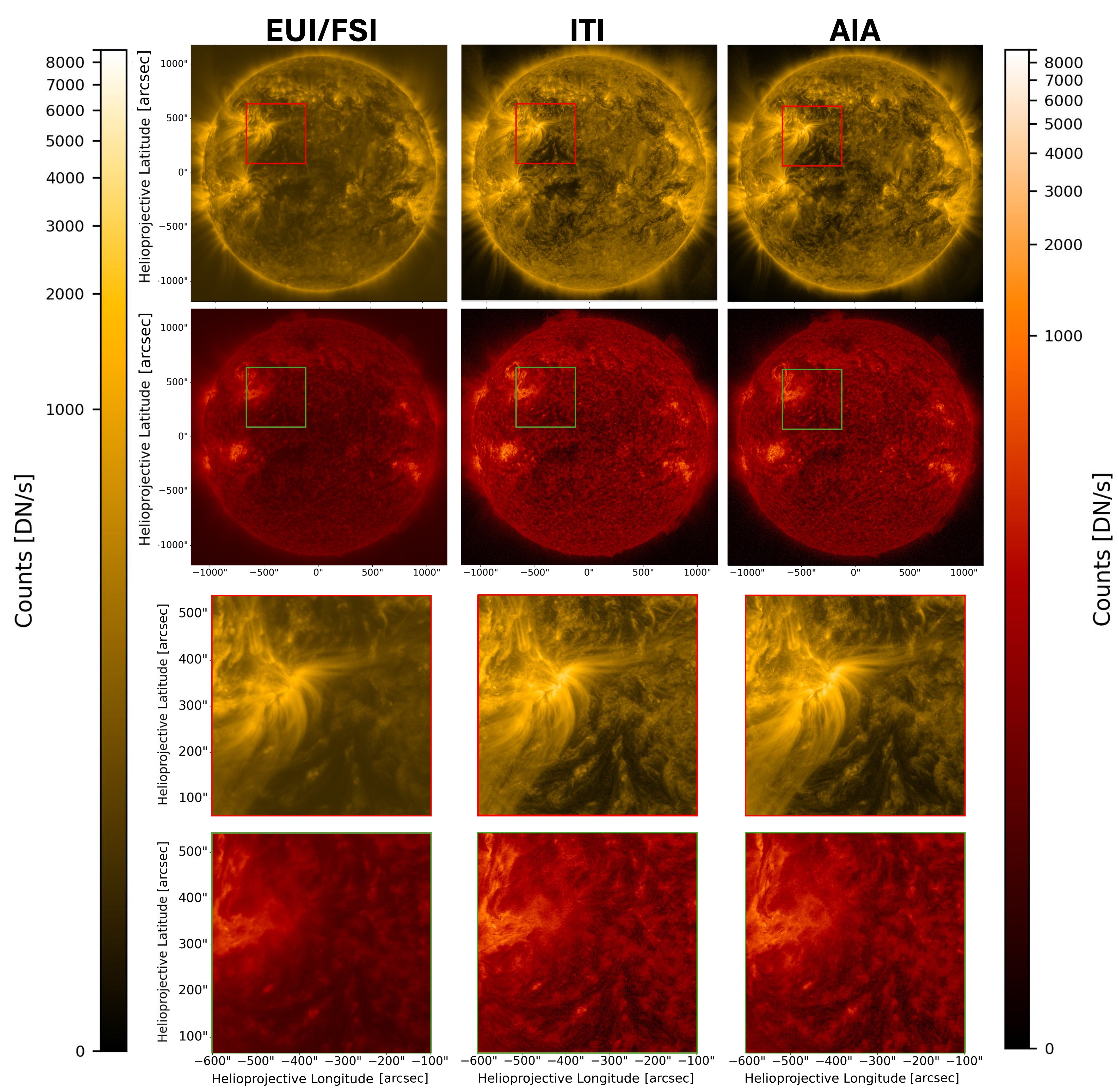}
   \caption{Comparison of the baseline calibration, ITI and the reference AIA observation from March 20 2024 at 16:00 UT. The left column shows the baseline calibrated FSI observations in 171~{\AA} and 304~{\AA}. The second column corresponds to our ITI translated observations and the third row shows the reference AIA observations. The smaller cutouts shown in the lower rows correspond to the green and red squares in the full Sun observations.}
\label{fig:FSI-ITI-AIA}
\end{figure*}

\subsection{Baseline calibration}
\label{subsec:base_calibration}
In order to evaluate the performance of our model we compare our ITI calibrations to a baseline calibration. Here, we calculate the mean and standard deviation of the FSI and AIA observations over the training set, to adjust the counts, given in dynamical number per second (DN/s) of FSI to AIA. For AIA, the mean and standard deviation are calculated over the time range May 2011 to April 2024, and for FSI from January 2022 to February 2023. These statistics are computed using the original intensity values (DN/s), without applying a logarithmic function. The baseline calibrated FSI observations are then calculated as:
\begin{equation}
\label{eq:basecalib}
    \text{FSI}_{cal,i}=[(\text{FSI}_i - \mu_{\text{FSI}}) / \sigma_{\text{FSI}}] * \sigma_{\text{AIA}} + \mu_{\text{AIA}},
\end{equation}
where $i$ corresponds to the $i$-th observation and $\mu$ and $\sigma$ indicates the mean and standard deviation calculated over the training set for each instrument. Since this calibration involves instruments observing the same scene, the observations are temporally aligned for evaluation to approximate the observation of the same scene by both spacecraft. Same as for the ITI calibrations, the degradation correction is also applied to the baseline calibrated FSI observations.

\subsection{Quality metric}
\label{subsec:qualmet}
We use the Fr\'{e}chet Inception Distance (FID; \citealp{Heusel2017}) to estimate the ability of our model to provide homogenized and intercalibrated data series. The FID evaluates the perceptual quality between two data sets, which do not need to be aligned on the pixel level. Instead of comparing individual images, the FID evaluates mean and covariances of a set of images and compares it to a ground truth data set. We calculate the FID between the baseline-calibrated FSI observations and the AIA observations. We refer to this as our "baseline". We then compare the FID of the baseline with the FID calculated between the ITI-translated observations and the AIA observations.

\subsection{Light curve comparison}
\label{subsec:lightcurve}
We evaluate the image calibration capability of ITI by comparing the light curves of SDO/AIA, the SolO/FSI baseline calibrated light curves and ITI. To this aim we compute the mean intensity per image and plot it as a function of time. Since SolO does not orbit the Sun in geosynchronous orbit as SDO does (Earth perspective), we adjust the timing of the FSI observations to match the orbital phase timing of the AIA observations. We therefore calculate the difference in longitude between the two spacecraft using the fact that the Sun completes one full synodic rotation, approximately every 27 days. This allows us to convert angular separations into time differences.
\begin{equation}
    t_{shift} = \frac{27}{360} \cdot \Delta \phi
\end{equation}
Here $\Delta \phi$ corresponds to the separation angle between SolO and SDO based on the Carrington rotation. Note that this is an approximation, which assumes rigid rotation and that the Sun is not significantly changing over up to one full rotation. This is not the case in reality. However, it is the only option to align the observations and to adjust the ITI and FSI observations and allows for an approximation of the intensity estimation at the time of the SDO/AIA observations, enabling a basic comparison of the individual instruments.

\begin{figure*}[h!tbp]
   \centering
   \includegraphics[scale=0.06]{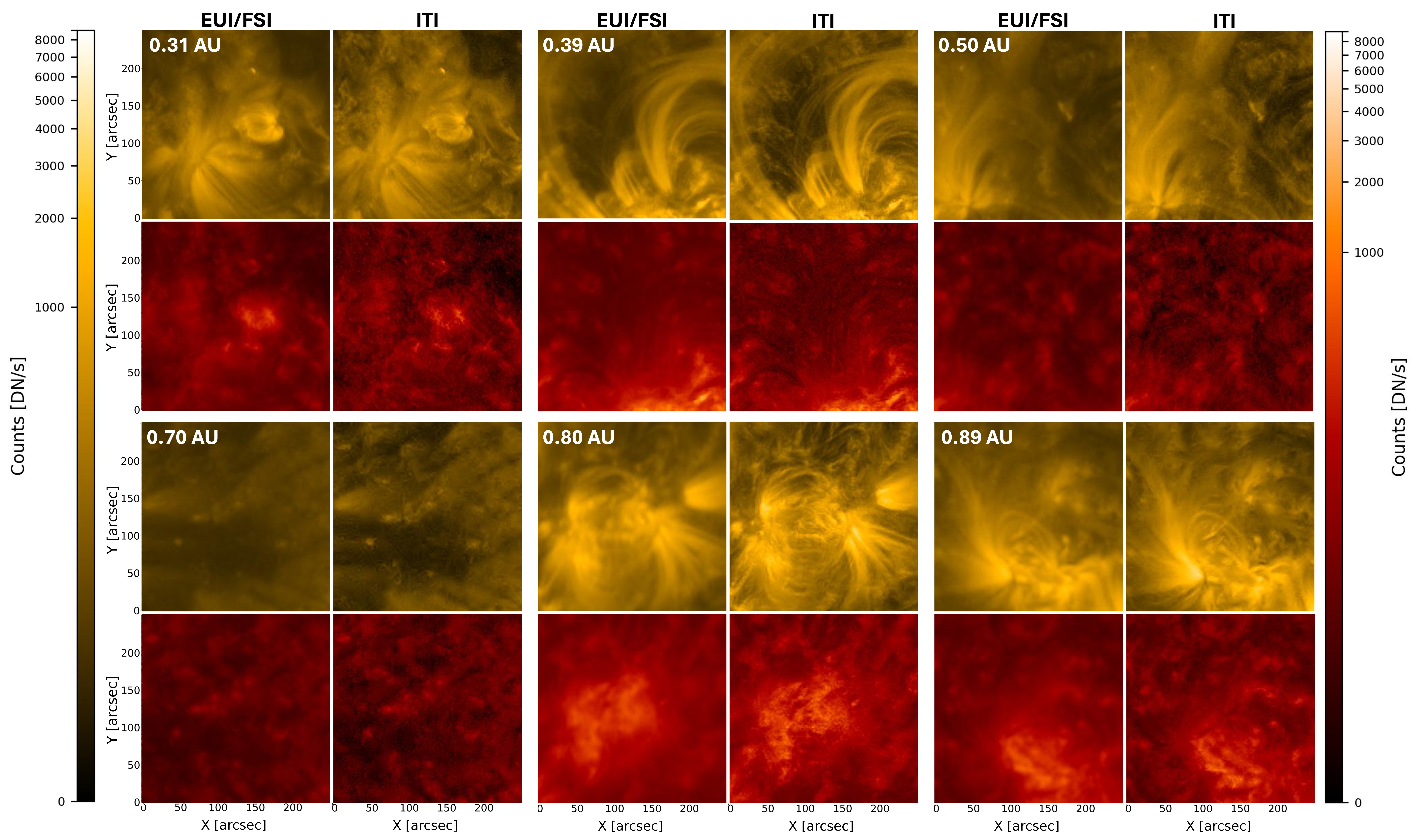}
   \caption{Comparison of the baseline calibrated FSI observations with our ITI calibrations for six different distances of SolO to the Sun. Each quad plot correspond to a certain distance, where the left column correspond to the EUI/FSI observation in 174~{\AA} and 304~{\AA} and the right column to the corresponding ITI calibration.}
\label{fig:distance}
\end{figure*}

\section{Results}
\label{sec:results}
The training for our ITI model is the same as described in \citet{Jarolim2025}. First, AIA observations of domain B are used to generate FSI observations in domain A using generator BA. A second neural network, generator AB, is trained to invert this translation by reconstructing AIA observations from the generated FSI observations. A discriminator network is used to enforce the domain translation by ensuring the generated observations match those from the target domain.
To evaluate the performance of the model, we use the test sets as described in Sect.\,\ref{sec:datasets} and summarized in Tab.\,\ref{table:dataset}. In this study, we evaluate and compare image distributions and light curves with a baseline calibration and we show a quantitative as well as a qualitative comparison where SolO was Earth aligned, for variable distances of SolO and for a shared FOV of SolO with SDO. 

\subsection{Quantitative comparison}
\label{subsec:quantcomp}
In  Fig.\,\ref{fig:histogram} we compare the logarithmic intensity distribution of the images from AIA (blue), the baseline calibrated FSI (green) as described in Sect.\,\ref{subsec:base_calibration} and ITI (red) for both channels. The plot shows the sum of the histograms over test set 1 for the 171~{\AA} (left) and the 304~{\AA} (right) filters on a logarithmic scale. The distributions show that for both channels, ITI matches the AIA distribution closer than the baseline calibrated FSI observations. We also checked the intensity distributions separately for the on-disk and off-limb pixels. Those are shown in the Appendix (Fig.\,\ref{fig:hist_comp_off_disk}). One can see that ITI matches the AIA distributions for both cases.

\begin{table}[h!tbp]
\caption{Comparison of the FID quality metric between the baseline and our ITI model and the MAE metric for the light curve evaluation in Fig.\,\ref{fig:lightcurve}.}             
\label{table:qualitymetrics}      
\centering                          
\begin{tabular}{c | c c c c | c c c}     
\hline\hline               
& & FID & & & & MAE & \\ 
& Baseline & & ITI & & Baseline & & ITI \\   
\hline
& & & & & & & \\
171 & 9.9 & & 3.5 & & 14.4 & & 0.6\\
& & & & & & & \\
304 & 11.9 & & 4.0 & & 2.9 & & 1.6 \\
\hline                           
\end{tabular}
\tablefoot{The comparison for the FID metrics is made using only test set 1 (Earth-aligned), whereas the MAE metric is calculated over the entire training set and test set 4 (Lightcurve)}.
\end{table}

We calculate the FID quality metric between the baseline calibrated FSI observation and the reference AIA observations, which we refer to our baseline and compare it with the FID, calculated between the ITI and the reference AIA observations. Table \ref{table:qualitymetrics} shows the perceptual quality assessment using the FID quality metric for the 171~{\AA} and 304~{\AA} channels, evaluated on test set 1 (Earth aligned). Lower FID values correspond to higher perceptual quality. With an FID of 3.5 for the 171 Å channel, ITI significantly outperforms the baseline calibration method, which has an FID of 9.9. Similar results are observed for the 304 Å channel, as shown in Tab.\,\ref{table:qualitymetrics}.

In Fig.\,\ref{fig:lightcurve} we evaluate the light curves, by calculating the mean of each observed image and plotting it as a function of time. To ensure a statistically significant data set for this comparison, the evaluation is performed on the entire FSI data set, including the training data and test set 4 (Lightcurve) until the end of January 2025. The FSI observations are baseline calibrated, where we shift the lightcurve to the mean and standard deviation of the AIA light curve according to Eq.\,\ref{eq:basecalib}. The mean and standard deviations are estimated over the entire training set for both FSI (2022--2023) and AIA (2011--2024). The first panel in Fig.\,\ref{fig:lightcurve}a shows the time-shifted light curve in the 171~{\AA} channel and the second panel the 304~{\AA} channel. The blue shaded area along the AIA light curve corresponds to a $\pm1\,\sigma$ error range. The gray shaded area represents the longitudinal position of SolO outside of the FOV of SDO ranging from $113^\circ$ to $248^\circ$. This is illustrated in panel b, where we indicate the related spacecraft constellations. We adjust the ITI and the baseline calibrated FSI observations according to the longitudinal difference of SolO and with respect to SDO to allow for a more consistent comparison as described in Sect.\,\ref{subsec:lightcurve}. For large separation angles, as shown in the third panel in Fig.\,\ref{fig:lightcurve}a, the results are more difficult to interpret since during this time SolO observed on the solar far-side. The dashed red line in Fig.\,\ref{fig:lightcurve}a indicates zero difference between the two spacecraft in longitude. The image calibration demonstrates robust performance across the entire period shown, with closer resemblance to the AIA ground truth than the baseline. Additionally, we evaluate the mean absolute error (MAE) of the baseline and ITI to the reference AIA light curve. ITI achieves a value of 0.6 for the 171~{\AA}, outperforming the baseline calibration with an MAE value of 14.4. Table\,\ref{table:qualitymetrics} list the MAE and FID values for both wavelengths. 

\subsection{Qualitative comparison}
\label{subsec:qualcomp}
Besides image calibration, ITI allows for simultaneous image enhancement to match the reference distribution as close as possible. This has already been shown for EUV instruments in previous applications of ITI i.e., for SOHO and STEREO \citep{Jarolim2025}. In Fig.\,\ref{fig:FSI-ITI-AIA} we show an example observation from test set 1 from 20 March 2024 at 16:00 UT. The distance of SolO to the Sun is $\Delta \text{D}=0.43$ AU and the difference in longitude and latitude between the SolO and SDO spacecraft is $\Delta \text{LON}=0.76^\circ$ and $\Delta \text{LAT}=0.67^\circ$, respectively. Figure\,\ref{fig:FSI-ITI-AIA} visualizes the full Sun observations as well as smaller cutouts. The cutouts are selected at random, with the aim of capturing both an active region and a quiet region. The first column corresponds to the baseline calibrated FSI observations, the second column to our ITI translations and the third column to the reference AIA observations. The first row shows the 171~{\AA} channel and the second row the 304~{\AA} channel. The full Sun observations show that our ITI translations show close resemblance to the reference AIA observations on the global scale. The zoomed-in region indicates that enhanced features are visually similar to structures observed by AIA on smaller scales. Where the baseline calibrated FSI observation looks more smooth with less contrast the ITI translated observation appears more sharp and the intensities matching the reference AIA observation closer. Similarly, in the 304~{\AA} channel, the loop like structure as well as the quiet Sun regions are represented sharper in ITI as compared to the baseline calibrated FSI cutouts. Note, even when Solar Orbiter is aligned with the Sun–Earth line, a persistent latitudinal offset between the two spacecraft introduces enhanced errors in pixel-based metrics and intensity ratio images.

\begin{figure}[h!tbp]
   \centering
   \includegraphics[scale=0.2]{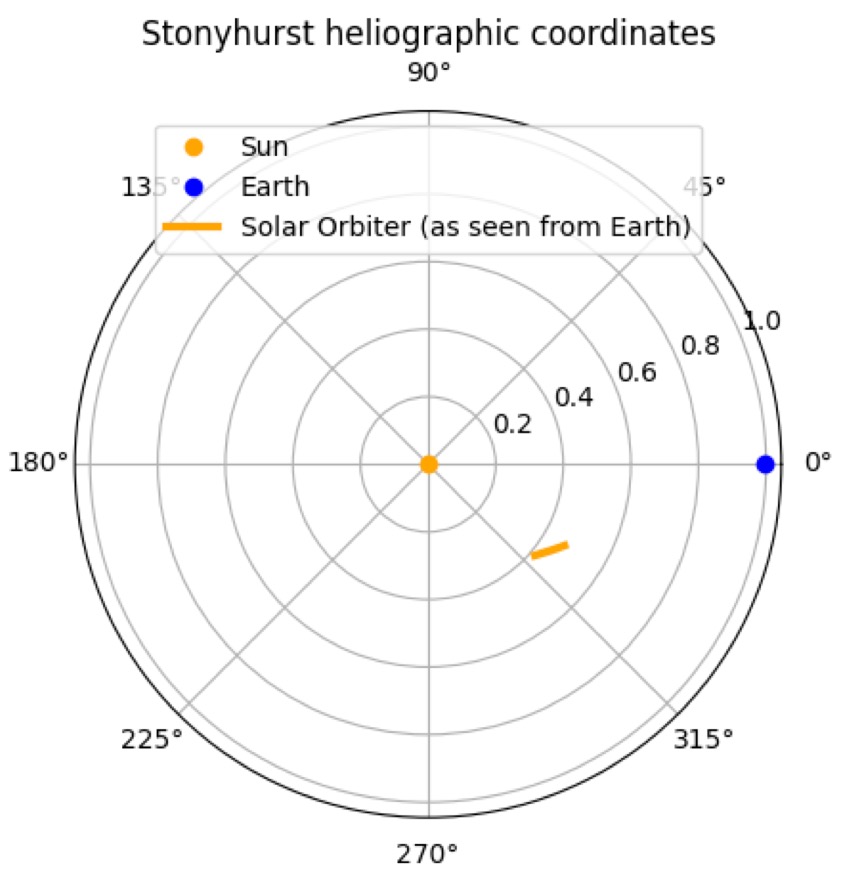}
   \caption{Solar Orbiter spacecraft position for the multi-viewpoint test set.}
\label{fig:mv_position}
\end{figure}

\subsection{Distance comparison}
\label{subsec:distance}

Due to the varying orbit of SolO, the spatial resolution of the observation also changes. For consistency, we resize and scale all observation to 1.1 solar radii and a resolution of 1024~$\times$~1024 pixels.
To see how the distance influences the ITI translations we test the ITI tool on test set 2 (Variable Distance) covering observations from 0.31 AU to 0.89 AU. The results are illustrated in Fig.\,\ref{fig:distance}. We show six observations, each for a particular distance from SolO to the Sun. The observations are cutouts with a size of $250''\times250''$. Each quad plot visualizes the baseline calibrated FSI observation in the left column for 174~{\AA} and 304~{\AA} channel and the right panel the corresponding ITI translations. We show observation at 0.31 AU, 0.39 AU, 0.50 AU, 0.70 AU, 0.80 AU and 0.89 AU (from top left to bottom right). 

It can be seen from the examples in Fig.\,\ref{fig:distance} how the spatial resolution decreases due to the radial distance of the SolO spacecraft to the Sun. For the first row (0.31 AU to 0.50 AU), the difference between FSI and ITI is dominated by the intensity calibration aspect to AIA, which shows a similar quality in terms of sharpness of structures, but with the intensities adapted to AIA. The second row (0.70 AU to 0.89 AU) shows that the original FSI observations have a lower spatial resolution. The ITI translated observations show an increase in contrast that matches the AIA observations. However, as the distance between Solo and the Sun increases, ITI can produce artifacts on small scales.

\subsection{Multi-viewpoint application}
\label{subsec:multiview}

\begin{figure*}[h!tbp]
   \centering
   \includegraphics[scale=0.13]{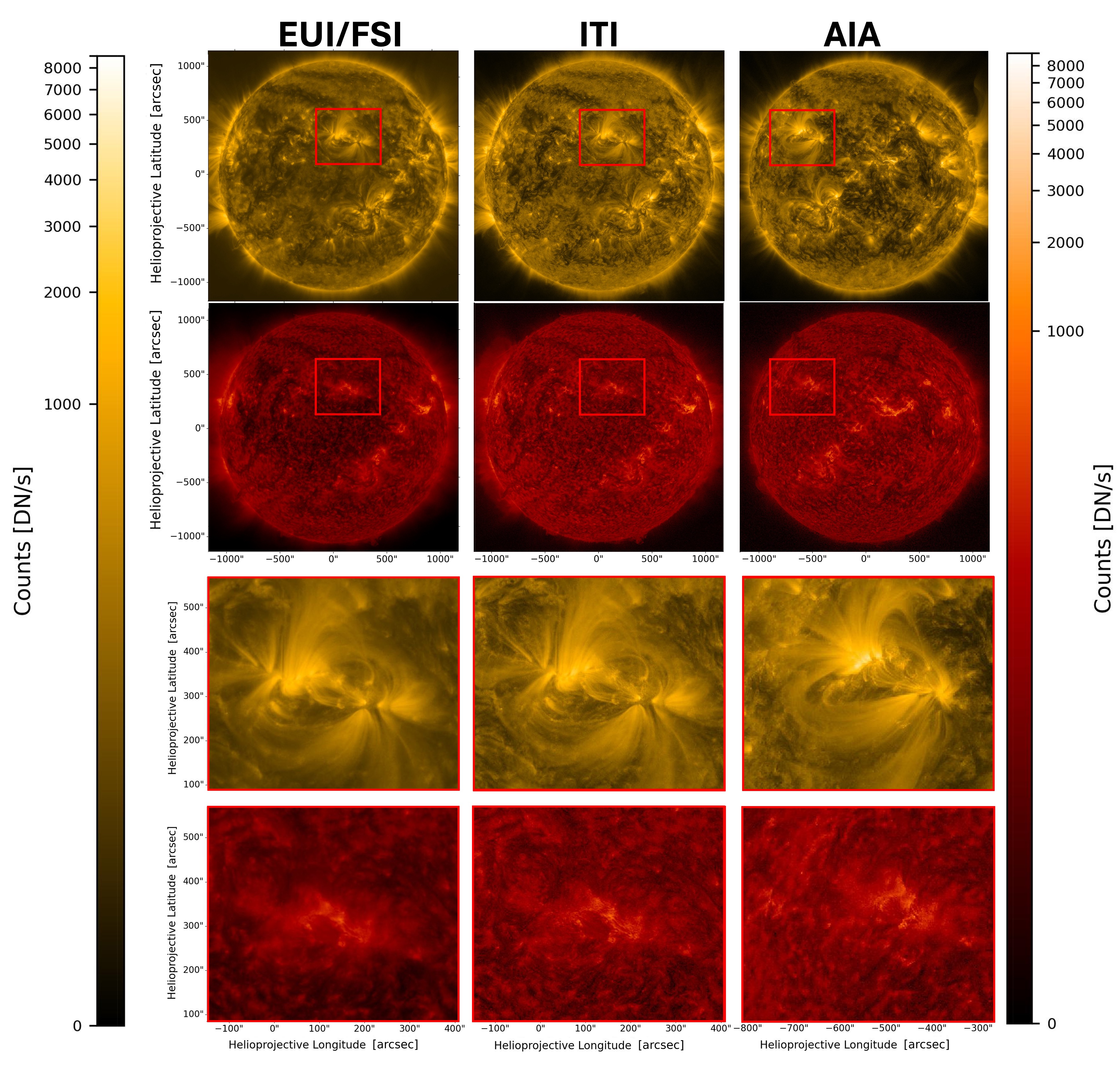}
   \caption{Multi-viewpoint comparison between SolO/EUI/FSI and SDO/AIA. The left column shows the baseline calibrated FSI observations, the middle column the ITI translated and the right column the AIA observations. The top row shows the 171~{\AA} and the bottom row the 304~{\AA} channel. The smaller cutouts correspond to the red rectangles in the full Sun observations.}
\label{fig:multi-viewpoint}
\end{figure*}

With the highly elliptical orbit of SolO, we can use ITI to provide joint data products with SDO/AIA and study the Sun from two different vantage points. We therefore use the third test set where SolO was not Earth aligned but part of its FOV overlap with the FOV of SDO from Earth perspective. The test set spans from 21 October 2023 to 24 October 2023 (test set 3). The spacecraft position of SolO (orange) and SDO (blue) is shown in Fig.\,\ref{fig:mv_position}. The separation angle of the two spacecraft in longitude across the test set ranges from $40^\circ$ to $33^\circ$ where SolO is west from SDO in longitude. 

Figure\,\ref{fig:multi-viewpoint} shows an example result of the test set from the 21 October 2023 at 00:00 UT. Same as in Fig.\,\ref{fig:FSI-ITI-AIA} we show the full Sun observations with smaller cutouts of an active region. The first column corresponds to the baseline calibrated FSI observations, the second column to the ITI translations and the third column to the AIA observations. The first and second row show the observations in 171~{\AA} and 304~{\AA}, respectively. For the zoomed-in views we cut out an active region located close to the disk center from SolOs perspective. The same active region is located on the eastern limb in the SDO/AIA perspective. A visual comparison on the global scale indicates that ITI shows a closer resemblance in contrast and quality to AIA compared to the baseline calibrated FSI, although they do not share the same FOV. Additionally, the quality of the zoomed in regions is also higher showing sharper contours for both 171~{\AA} and 304~{\AA} in ITI. The contrast difference between FSI, ITI and AIA in both channels can be explained due to a different line-of-sight integration since the location of the active region for SolO is on the disc center and for AIA closer to the limb. We point out that these intercalibrated data products cannot be quantified because of the large separation angle of the two spacecraft.

\section{Discussion and conclusion}
\label{sec:conclusion}
We applied the ITI framework from \cite{Jarolim2025} to calibrate the FSI instrument onboard SolO with the AIA instrument onboard SDO. The unpaired image translation approach allows to calibrate both instruments, which would not be achievable with paired image translation due to the sparse overlap of observations \citep{Isola2017}. 

We compare our results against a standard baseline calibration as described in Sect. \ref{subsec:base_calibration}. Comparing the light curves in Fig.\,\ref{fig:lightcurve} we see that ITI provides an improved calibration of the intensities and therefore outperforms the baseline calibration. This is also quantified in Tab.\,\ref{table:qualitymetrics} where the MAE of ITI shows better image calibrations compared to the baseline. The comparison of the image distributions in Fig.\,\ref{fig:histogram} and the FID quality metric in Tab.\,\ref{table:qualitymetrics} also confirms this for perceptual quality for both channels, where lower FID values correspond to higher perceptual quality. Comparing the calibrations of ITI for different distances of SolO to the Sun in Fig.\,\ref{fig:distance}, we see that for close in observations (0.3 - 0.5 AU) the intensity adjustment dominates, as the spatial resolution of the original FSI observation is high. In contrast, the spatial resolution of FSI decreases with distance, resulting in more blurred solar features. ITI is capable of adjusting the intensities but also enhances the observations showing more details on the smaller scales. However, this can also lead to small scale artifacts in the ITI calibrations for both the 171~{\AA} and 304~{\AA} filters. Consequently, ITI provides homogeneous data series and can be used for both global and small scale studies for distances of SolO up to 0.5 AU from the Sun due to the high spatial resolution of the FSI instrument. For distances greater than 0.5 AU, ITI can be used for studies on global scales. These distances should not be used for the study of small scale features, as artifacts can appear at smaller scales. In addition, off-pointing SOOP observations cannot be used for ITI image calibration, as these observations show intensity variations in different regions on the solar disk that ITI cannot account for.

Figure\,\ref{fig:lightcurve} illustrates the importance of image calibration, as the Sun shows substantial variability over solar rotation and solar cycle time scales. The limitation here is that we cannot quantify these changes for ITI over a longer period, primarily due to the different orbital paths of SolO and SDO. Multi-viewpoint observations are therefore essential for accurate quantification.
Nevertheless, based on light assumptions, we show that ITI performs well on average for image calibration from different viewpoints as well as on the solar far-side. As a result, ITI not only provides reliable calibrations across instruments, but also enables pre- and post- estimations of solar EUV irradiance, with intensities adapted to AIA, as the Sun rotates in and out of Earth's FOV \citep{woods1998, Lilensten2008}. 

In conclusion, ITI can effectively overcome limitations of standard image calibration methods, accounting for instrumental effects, filter characteristics, and quality differences by leveraging the fully data driven approach of high-quality data sets \citep{Hamada2020}. The improved ITI calibrated observations enable more consistent solar cycle studies such as solar variability investigations where multi-instrument data need to be combined to a homogeneous data series \citep{ermolli2013}. With the out-of-ecliptic transition of SolO in 2025, ITI can be used to obtain calibrated data products that will allow to study the Sun from multiple vantage points and outside the ecliptic plane \citep{saqri2022, purkhart2024}. Additionally, the results demonstrate that the calibrated ITI data products can be used for active regions and coronal hole evolution and detection (e.g., \citealp{Jarolim2021, Delouille2018}).

The next step involves making the ITI calibrated data set available to the scientific community. The code used for the translations is already publicly accessible. Additionally, we aim to constantly update and integrate new data to the framework from future space missions i.e. from the Joint EUV coronal Diagnostic Investigation (JEDI) imager onboard the upcoming Vigil mission \citep{Palomba2022} in order to provide a large database of homogenized data products.

\section*{Code availability}
The code is publicly available in the following repository: \url{https://github.com/spaceml-org/InstrumentToInstrument}. In addition, we provide framework documentation with example Jupyter notebooks on our ReadTheDocs website: \url{https://iti-documentation.readthedocs.io}. All ITI models are publicly available at \url{https://huggingface.co/SpaceML/ITI}.

\begin{acknowledgements}
This research has received financial support from NASA award 22-MDRAIT22-0018 (No. 80NSSC23K1045) and managed by Trillium Technologies, Inc. The research was in part sponsored by the DynaSun project and has thus received funding under the Horizon Europe programme of the European Union under grant agreement (no. 101131534). Views and opinions expressed are however those of the author(s) only and do not necessarily reflect those of the European Union and therefore the European Union cannot be held responsible for them. CS was supported by the University of Graz EST (European Solar Telescope) program and the Guest Investigator Program of the Royal Observatory of Belgium (ROB). RJ was supported by the NASA Jack-Eddy Fellowship. LD was supported by the PRODEX Programme of the European Space Agency (ESA) under contract number 4000136424. We thankfully acknowledge David Berghmans for providing the Solar Orbiter data and the valuable comments to the paper.
Solar Orbiter is a space mission of international collaboration between ESA and NASA, operated by ESA. The EUI instrument was built by CSL, IAS, MPS, MSSL/UCL, PMOD/WRC, ROB, LCF/IO with funding from the Belgian Federal Science Policy Office (BELSPO/PRODEX PEA 4000112292 and 4000134088); the Centre National d’Etudes Spatiales (CNES); the UK Space Agency (UKSA); the Bundesministerium für Wirtschaft und Energie (BMWi) through the Deutsches Zentrum für Luft- und Raumfahrt (DLR); and the Swiss Space Office (SSO).
This research has made use of AstroPy \citep{astropy2022}, SunPy \citep{sunpy2020} and PyTorch \citep{pytorch2017}. 
\end{acknowledgements}

\bibliographystyle{aa}
\bibliography{bib}

\begin{thebibliography}{50}
\expandafter\ifx\csname natexlab\endcsname\relax\def\natexlab#1{#1}\fi

\bibitem[{Aschwanden {et~al.}(2008)Aschwanden, W{\"u}lser, Nitta, \& Lemen}]{Aschwanden2008_triang}
Aschwanden, M.~J., W{\"u}lser, J.-P., Nitta, N.~V., \& Lemen, J.~R. 2008, The Astrophysical Journal, 679, 827

\bibitem[{{Astropy Collaboration} {et~al.}(2022){Astropy Collaboration}, {Price-Whelan}, {Lim}, {Earl}, {Starkman}, {Bradley}, {Shupe}, {Patil}, {Corrales}, {Brasseur}, {N{"o}the}, {Donath}, {Tollerud}, {Morris}, {Ginsburg}, {Vaher}, {Weaver}, {Tocknell}, {Jamieson}, {van Kerkwijk}, {Robitaille}, {Merry}, {Bachetti}, {G{"u}nther}, {Aldcroft}, {Alvarado-Montes}, {Archibald}, {B{'o}di}, {Bapat}, {Barentsen}, {Baz{'a}n}, {Biswas}, {Boquien}, {Burke}, {Cara}, {Cara}, {Conroy}, {Conseil}, {Craig}, {Cross}, {Cruz}, {D'Eugenio}, {Dencheva}, {Devillepoix}, {Dietrich}, {Eigenbrot}, {Erben}, {Ferreira}, {Foreman-Mackey}, {Fox}, {Freij}, {Garg}, {Geda}, {Glattly}, {Gondhalekar}, {Gordon}, {Grant}, {Greenfield}, {Groener}, {Guest}, {Gurovich}, {Handberg}, {Hart}, {Hatfield-Dodds}, {Homeier}, {Hosseinzadeh}, {Jenness}, {Jones}, {Joseph}, {Kalmbach}, {Karamehmetoglu}, {Ka{l}uszy{'n}ski}, {Kelley}, {Kern}, {Kerzendorf}, {Koch}, {Kulumani}, {Lee}, {Ly}, {Ma}, {MacBride}, {Maljaars}, {Muna}, {Murphy}, {Norman}, {O'Steen},
  {Oman}, {Pacifici}, {Pascual}, {Pascual-Granado}, {Patil}, {Perren}, {Pickering}, {Rastogi}, {Roulston}, {Ryan}, {Rykoff}, {Sabater}, {Sakurikar}, {Salgado}, {Sanghi}, {Saunders}, {Savchenko}, {Schwardt}, {Seifert-Eckert}, {Shih}, {Jain}, {Shukla}, {Sick}, {Simpson}, {Singanamalla}, {Singer}, {Singhal}, {Sinha}, {Sip{H{o}}cz}, {Spitler}, {Stansby}, {Streicher}, {{{S}}umak}, {Swinbank}, {Taranu}, {Tewary}, {Tremblay}, {Val-Borro}, {Van Kooten}, {Vasovi{'c}}, {Verma}, {de Miranda Cardoso}, {Williams}, {Wilson}, {Winkel}, {Wood-Vasey}, {Xue}, {Yoachim}, {Zhang}, {Zonca}, \& {Astropy Project Contributors}}]{astropy2022}
{Astropy Collaboration}, {Price-Whelan}, A.~M., {Lim}, P.~L., {et~al.} 2022, \apj, 935, 167

\bibitem[{Barnes {et~al.}(2020)Barnes, m.~Cheung, Bobra, Boerner, Chintzoglou, Leonard, Mumford, Padmanabhan, Shih, Shirman, Stansby, \& Wright}]{Barnes2020}
Barnes, W.~T., m.~Cheung, M.~C., Bobra, M.~G., {et~al.} 2020, Journal of Open Source Software, 5, 2801

\bibitem[{Berghmans {et~al.}(2021)Berghmans, Auchère, Long, Soubrié, Mierla, Zhukov, Schühle, Antolin, Harra, Parenti, Podladchikova, Aznar~Cuadrado, Buchlin, Dolla, Verbeeck, Gissot, Teriaca, Haberreiter, Katsiyannis, Rodriguez, Kraaikamp, Smith, Stegen, Rochus, Halain, Jacques, Thompson, \& Inhester}]{Berghmans2021}
Berghmans, D., Auchère, F., Long, D.~M., {et~al.} 2021, A\&A, 656, L4

\bibitem[{Caplan {et~al.}(2016)Caplan, Downs, \& Linker}]{Caplan2016}
Caplan, R.~M., Downs, C., \& Linker, J.~A. 2016, The Astrophysical Journal, 823, 53

\bibitem[{Chitta {et~al.}(2023)Chitta, Zhukov, Berghmans, Peter, Parenti, Mandal, Aznar~Cuadrado, Sch{\"u}hle, Teriaca, Auch{\`e}re, {et~al.}}]{Chitta2023}
Chitta, L.~P., Zhukov, A.~N., Berghmans, D., {et~al.} 2023, Science, 381, 867

\bibitem[{Collier {et~al.}(2024)Collier, Hayes, Purkhart, Krucker, Ryan, Polito, Veronig, Harra, Berghmans, Kraaikamp, Dominique, Dolla, \& Verbeeck}]{Collier2024}
Collier, H., Hayes, L., Purkhart, S., {et~al.} 2024, Astronomy \& Astrophysics, 692

\bibitem[{Delouille {et~al.}(2018)Delouille, Hofmeister, Reiss, Mampaey, Temmer, \& Veronig}]{Delouille2018}
Delouille, V., Hofmeister, S.~J., Reiss, M.~A., {et~al.} 2018, in Machine learning techniques for space weather (Elsevier), 365--395

\bibitem[{Domingo {et~al.}(1995)Domingo, Fleck, \& Poland}]{Domingo1995}
Domingo, V., Fleck, B., \& Poland, A.~I. 1995, \solphys, 162, 1

\bibitem[{Dos~Santos {et~al.}(2021)Dos~Santos, Bose, Salvatelli, Neuberg, Cheung, Janvier, Jin, Gal, Boerner, \& Baydin}]{DosSantos2021}
Dos~Santos, L. F.~G., Bose, S., Salvatelli, V., {et~al.} 2021, A\&A, 648, A53

\bibitem[{Driesman {et~al.}(2008)Driesman, Hynes, \& Cancro}]{Driesman2008}
Driesman, A., Hynes, S., \& Cancro, G. 2008, \ssr, 136, 17

\bibitem[{Díaz~Baso \& Asensio~Ramos(2018)}]{Baso2018}
Díaz~Baso, C.~J. \& Asensio~Ramos, A. 2018, A\&A, 614, A5

\bibitem[{Díaz~Baso {et~al.}(2019)Díaz~Baso, de~la Cruz~Rodríguez, \& Danilovic}]{Baso2019}
Díaz~Baso, C.~J., de~la Cruz~Rodríguez, J., \& Danilovic, S. 2019, A\&A, 629, A99

\bibitem[{Ermolli {et~al.}(2013)Ermolli, Matthes, Dudok~de Wit, Krivova, Tourpali, Weber, Unruh, Gray, Langematz, Pilewskie, {et~al.}}]{ermolli2013}
Ermolli, I., Matthes, K., Dudok~de Wit, T., {et~al.} 2013, Atmospheric Chemistry and Physics, 13, 3945

\bibitem[{Goodfellow {et~al.}(2020)Goodfellow, Pouget-Abadie, Mirza, Xu, Warde-Farley, Ozair, Courville, \& Bengio}]{Goodfellow2014}
Goodfellow, I., Pouget-Abadie, J., Mirza, M., {et~al.} 2020, Commun. ACM, 63, 139–144

\bibitem[{Hamada {et~al.}(2020)Hamada, Asikainen, \& Mursula}]{Hamada2020}
Hamada, A., Asikainen, T., \& Mursula, K. 2020, \solphys, 295, 2

\bibitem[{Heusel {et~al.}(2017)Heusel, Ramsauer, Unterthiner, Nessler, \& Hochreiter}]{Heusel2017}
Heusel, M., Ramsauer, H., Unterthiner, T., Nessler, B., \& Hochreiter, S. 2017, in Proceedings of the 31st International Conference on Neural Information Processing Systems, NIPS'17 (Red Hook, NY, USA: Curran Associates Inc.), 6629–6640

\bibitem[{Isola {et~al.}(2017)Isola, Zhu, Zhou, \& Efros}]{Isola2017}
Isola, P., Zhu, J.-Y., Zhou, T., \& Efros, A.~A. 2017, in 2017 IEEE Conference on Computer Vision and Pattern Recognition (CVPR), 5967--5976

\bibitem[{Jarolim {et~al.}(2024)Jarolim, Tremblay, Muñoz-Jaramillo, Bintsi, Jungbluth, Santos, Vourlidas, Mason, Sundaresan, Downs, \& Caplan}]{Jarolim2024}
Jarolim, R., Tremblay, B., Muñoz-Jaramillo, A., {et~al.} 2024, The Astrophysical Journal Letters, 961, L31

\bibitem[{Jarolim {et~al.}(2025)Jarolim, Veronig, P{\"o}tzi, \& Podladchikova}]{Jarolim2025}
Jarolim, R., Veronig, A., P{\"o}tzi, W., \& Podladchikova, T. 2025, Nature Communications, 16, 3157

\bibitem[{Jarolim {et~al.}(2021)Jarolim, Veronig, Hofmeister, Heinemann, Temmer, Podladchikova, \& Dissauer}]{Jarolim2021}
Jarolim, R., Veronig, A.~M., Hofmeister, S., {et~al.} 2021, A\&A, 652, A13

\bibitem[{Kim {et~al.}(2019)Kim, Park, Lee, Moon, Bae, Lim, Jang, Kim, Cho, Choi, {et~al.}}]{Kim2019}
Kim, T., Park, E., Lee, H., {et~al.} 2019, Nature Astronomy, 3, 397

\bibitem[{Kraaikamp {et~al.}(2023)Kraaikamp, Gissot, Stegen, Mampaey, Verbeeck, Auch{\`e}re, \& Berghmans}]{euidatarelease6}
Kraaikamp, E., Gissot, S., Stegen, K., {et~al.} 2023, SolO/EUI Data Release 6.0 2023-01, https://doi.org/10.24414/z818-4163, published by Royal Observatory of Belgium (ROB)

\bibitem[{Lemen {et~al.}(2012)Lemen, Title, Akin, Boerner, Chou, Drake, Duncan, Edwards, Friedlaender, Heyman, Hurlburt, Katz, Kushner, Levay, Lindgren, Mathur, McFeaters, Mitchell, Rehse, \& Waltham}]{Lemen2011}
Lemen, J., Title, A., Akin, D., {et~al.} 2012, solphys, 275, 17

\bibitem[{Lilensten {et~al.}(2008)Lilensten, Dudok~de Wit, Kretzschmar, Amblard, Moussaoui, Aboudarham, \& Auch\`ere}]{Lilensten2008}
Lilensten, J., Dudok~de Wit, T., Kretzschmar, M., {et~al.} 2008, Annales Geophysicae, 26, 269

\bibitem[{Müller {et~al.}(2020)Müller, St.~Cyr, Zouganelis, Gilbert, Marsden, Nieves-Chinchilla, Antonucci, Auchère, Berghmans, Horbury, Howard, Krucker, Maksimovic, Owen, Rochus, Rodriguez-Pacheco, Romoli, Solanki, Bruno, Carlsson, Fludra, Harra, Hassler, Livi, Louarn, Peter, Schühle, Teriaca, del Toro~Iniesta, Wimmer-Schweingruber, Marsch, Velli, De~Groof, Walsh, \& Williams}]{Mueller2020}
Müller, D., St.~Cyr, O.~C., Zouganelis, I., {et~al.} 2020, A\&A, 642, A1

\bibitem[{Nistic{\'o} {et~al.}(2017)Nistic{\'o}, Polito, Nakariakov, \& Del~Zanna}]{Nistico2017}
Nistic{\'o}, G., Polito, V., Nakariakov, V.~M., \& Del~Zanna, G. 2017, A\&A, 600, A37

\bibitem[{O’Kane {et~al.}(2021)O’Kane, Green, Davies, Möstl, Hinterreiter, Freiherr~von Forstner, Weiss, Long, \& Amerstorfer}]{Kane2021}
O’Kane, J., Green, L.~M., Davies, E.~E., {et~al.} 2021, A\&A, 656, L6

\bibitem[{Palomba \& Luntama(2022)}]{Palomba2022}
Palomba, M. \& Luntama, J.-P. 2022, in 44th COSPAR Scientific Assembly. Held 16-24 July, Vol.~44, 3544

\bibitem[{Paszke {et~al.}(2017)Paszke, Gross, Chintala, Chanan, Yang, DeVito, Lin, Desmaison, Antiga, \& Lerer}]{pytorch2017}
Paszke, A., Gross, S., Chintala, S., {et~al.} 2017, in NIPS-W

\bibitem[{Pesnell {et~al.}(2012)Pesnell, Thompson, \& Chamberlin}]{Pesnell2012}
Pesnell, W., Thompson, B., \& Chamberlin, P. 2012, solphys, 275, 3

\bibitem[{Podladchikova {et~al.}(2024)Podladchikova, Jain, Veronig, Purkhart, Chikunova, Dissauer, \& Dumbovi{\'c}}]{podladchikova2024}
Podladchikova, T., Jain, S., Veronig, A.~M., {et~al.} 2024, Astronomy \& Astrophysics, 691, A344

\bibitem[{P{\"o}tzi {et~al.}(2021)P{\"o}tzi, Veronig, Jarolim, Rodr{\'\i}guez~G{\'o}mez, Podlachikova, Baumgartner, Freislich, \& Strutzmann}]{Potzi2021}
P{\"o}tzi, W., Veronig, A., Jarolim, R., {et~al.} 2021, Solar Physics, 296, 164

\bibitem[{Purkhart {et~al.}(2024)Purkhart, Veronig, Kliem, Jarolim, Dissauer, Dickson, Podladchikova, \& Krucker}]{purkhart2024}
Purkhart, S., Veronig, A.~M., Kliem, B., {et~al.} 2024, Astronomy \& Astrophysics, 689, A259

\bibitem[{Reiss {et~al.}(2021)Reiss, Muglach, Möstl, Arge, Bailey, Delouille, Garton, Hamada, Hofmeister, Illarionov, Jarolim, Kirk, Kosovichev, Krista, Lee, Lowder, MacNeice, Veronig, \& Team}]{Reiss2021}
Reiss, M.~A., Muglach, K., Möstl, C., {et~al.} 2021, The Astrophysical Journal, 913, 28

\bibitem[{Rochus {et~al.}(2020)Rochus, Auch{\`e}re, Berghmans, Harra, Schmutz, Sch{\"u}hle, Addison, Appourchaux, Aznar~Cuadrado, Baker, Barbay, Bates, BenMoussa, Bergmann, Beurthe, Borgo, Bonte, Bouzit, Bradley, B{\"u}chel, Buchlin, B{\"u}chner, Cab{\'e}, Cadiergues, Chaigneau, Chares, Choque~Cortez, Coker, Condamin, Coumar, Curdt, Cutler, Davies, Davison, Defise, Del~Zanna, Delmotte, Delouille, Dolla, Dumesnil, D{\"u}rig, Enge, Fran{\c{c}}ois, Fourmond, Gillis, Giordanengo, Gissot, Green, Guerreiro, Guilbaud, Gyo, Haberreiter, Hafiz, Hailey, Halain, Hansotte, Hecquet, Heerlein, Hellin, Hemsley, Hermans, Hervier, Hochedez, Houbrechts, Ihsan, Jacques, J{\'e}r{\^o}me, Jones, Kahle, Kennedy, Klaproth, Kolleck, Koller, Kotsialos, Kraaikamp, Langer, Lawrenson, Le~Clech', Lenaerts, Liebecq, Linder, Long, Mampaey, Markiewicz-Innes, Marquet, Marsch, Matthews, Mazy, Mazzoli, Meining, Meltchakov, Mercier, Meyer, Monecke, Monfort, Morinaud, Moron, Mountney, M{\"u}ller, Nicula, Parenti, Peter, Pfiffner, Philippon,
  Phillips, Plesseria, Pylyser, Rabecki, Ravet-Krill, Rebellato, Renotte, Rodriguez, Roose, Rosin, Rossi, Roth, Rouesnel, Roulliay, Rousseau, Ruane, Scanlan, Schlatter, Seaton, Silliman, Smit, Smith, Solanki, Spescha, Spencer, Stegen, Stockman, Szwec, Tamiatto, Tandy, Teriaca, Theobald, Tychon, van Driel-Gesztelyi, Verbeeck, Vial, Werner, West, Westwood, Wiegelmann, Willis, Winter, Zerr, Zhang, \& Zhukov}]{Rochus2020}
Rochus, P., Auch{\`e}re, F., Berghmans, D., {et~al.} 2020, \aap, 642, A8

\bibitem[{Ronneberger {et~al.}(2015)Ronneberger, Fischer, \& Brox}]{Ronneberger2015}
Ronneberger, O., Fischer, P., \& Brox, T. 2015, in Medical Image Computing and Computer-Assisted Intervention -- MICCAI 2015, ed. N.~Navab, J.~Hornegger, W.~M. Wells, \& A.~F. Frangi (Cham: Springer International Publishing), 234--241

\bibitem[{Saqri {et~al.}(2024)Saqri, Veronig, Battaglia, Dickson, Gary, \& Krucker}]{saqri2024}
Saqri, J., Veronig, A.~M., Battaglia, A.~F., {et~al.} 2024, Astronomy \& Astrophysics, 683, A41

\bibitem[{Saqri {et~al.}(2022)Saqri, Veronig, Warmuth, Dickson, Battaglia, Podladchikova, Xiao, Battaglia, Hurford, \& Krucker}]{saqri2022}
Saqri, J., Veronig, A.~M., Warmuth, A., {et~al.} 2022, Astronomy \& Astrophysics, 659, A52

\bibitem[{Schirninger {et~al.}(2024)Schirninger, Jarolim, Veronig, \& Kuckein}]{Schirninger2024}
Schirninger, C., Jarolim, R., Veronig, A.~M., \& Kuckein, C. 2024, Astronomy and astrophysics., 693, A6

\bibitem[{Schwanitz {et~al.}(2023)Schwanitz, Harra, Mandrini, Sterling, Raouafi, Mac~Cormack, Berghmans, Auch{\`e}re, Barczynski, Cuadrado, {et~al.}}]{Schwanitz2023}
Schwanitz, C., Harra, L., Mandrini, C.~H., {et~al.} 2023, Astronomy \& Astrophysics, 674, A219

\bibitem[{{The SunPy Community} {et~al.}(2020){The SunPy Community}, Barnes, Bobra, Christe, Freij, Hayes, Ireland, Mumford, Perez-Suarez, Ryan, Shih, Chanda, Glogowski, Hewett, Hughitt, Hill, Hiware, Inglis, Kirk, Konge, Mason, Maloney, Murray, Panda, Park, Pereira, Reardon, Savage, Sipőcz, Stansby, Jain, Taylor, Yadav, Rajul, \& Dang}]{sunpy2020}
{The SunPy Community}, Barnes, W.~T., Bobra, M.~G., {et~al.} 2020, ApJ, 890, 68

\bibitem[{Tsuneta {et~al.}(2008)Tsuneta, Ichimoto, Katsukawa, Nagata, Otsubo, Shimizu, Suematsu, Nakagiri, Noguchi, Tarbell, {et~al.}}]{Tsuneta2008}
Tsuneta, S., Ichimoto, K., Katsukawa, Y., {et~al.} 2008, Solar Physics, 249, 167

\bibitem[{Usoskin \& Mursula(2003)}]{Usoskin2003}
Usoskin, I. \& Mursula, K. 2003, Solar Physics, 218, 319

\bibitem[{Wang {et~al.}(2018)Wang, Liu, Zhu, Tao, Kautz, \& Catanzaro}]{Wang2018}
Wang, T.-C., Liu, M.-Y., Zhu, J.-Y., {et~al.} 2018, in 2018 IEEE/CVF Conference on Computer Vision and Pattern Recognition, 8798--8807

\bibitem[{Woods {et~al.}(2012)Woods, Eparvier, Hock, Jones, Woodraska, Judge, Didkovsky, Lean, Mariska, Warren, {et~al.}}]{Woods2012}
Woods, T.~N., Eparvier, F., Hock, R., {et~al.} 2012, The solar dynamics observatory, 115

\bibitem[{Woods {et~al.}(1998)Woods, Rottman, Bailey, Solomon, \& Worden}]{woods1998}
Woods, T.~N., Rottman, G.~J., Bailey, S.~M., Solomon, S.~C., \& Worden, J.~R. 1998, in Solar Electromagnetic Radiation Study for Solar Cycle 22: Proceedings of the SOLERS22 Workshop held at the National Solar Observatory, Sacramento Peak, Sunspot, New Mexico, USA, June 17--21, 1996, Springer, 133--146

\bibitem[{Youn {et~al.}(2025)Youn, Lee, Jeong, Lee, Park, \& Moon}]{youn2025}
Youn, J., Lee, H., Jeong, H.-J., {et~al.} 2025, Astronomy \& Astrophysics, 695, A125

\bibitem[{Zhu {et~al.}(2017)Zhu, Park, Isola, \& Efros}]{Zhu2017}
Zhu, J.-Y., Park, T., Isola, P., \& Efros, A.~A. 2017, in 2017 IEEE International Conference on Computer Vision (ICCV), 2242--2251

\bibitem[{{Zouganelis, I.} {et~al.}(2020){Zouganelis, I.}, {De Groof, A.}, {Walsh, A. P.}, {Williams, D. R.}, {Müller, D.}, {St Cyr, O. C.}, {Auchère, F.}, {Berghmans, D.}, {Fludra, A.}, {Horbury, T. S.}, {Howard, R. A.}, {Krucker, S.}, {Maksimovic, M.}, {Owen, C. J.}, {Rodríguez-Pacheco, J.}, {Romoli, M.}, {Solanki, S. K.}, {Watson, C.}, {Sanchez, L.}, {Lefort, J.}, {Osuna, P.}, {Gilbert, H. R.}, {Nieves-Chinchilla, T.}, {Abbo, L.}, {Alexandrova, O.}, {Anastasiadis, A.}, {Andretta, V.}, {Antonucci, E.}, {Appourchaux, T.}, {Aran, A.}, {Arge, C. N.}, {Aulanier, G.}, {Baker, D.}, {Bale, S. D.}, {Battaglia, M.}, {Bellot Rubio, L.}, {Bemporad, A.}, {Berthomier, M.}, {Bocchialini, K.}, {Bonnin, X.}, {Brun, A. S.}, {Bruno, R.}, {Buchlin, E.}, {Büchner, J.}, {Bucik, R.}, {Carcaboso, F.}, {Carr, R.}, {Carrasco-Blázquez, I.}, {Cecconi, B.}, {Cernuda Cangas, I.}, {Chen, C. H. K.}, {Chitta, L. P.}, {Chust, T.}, {Dalmasse, K.}, {D’Amicis, R.}, {Da Deppo, V.}, {De Marco, R.}, {Dolei, S.}, {Dolla, L.}, {Dudok de
  Wit, T.}, {van Driel-Gesztelyi, L.}, {Eastwood, J. P.}, {Espinosa Lara, F.}, {Etesi, L.}, {Fedorov, A.}, {Félix-Redondo, F.}, {Fineschi, S.}, {Fleck, B.}, {Fontaine, D.}, {Fox, N. J.}, {Gandorfer, A.}, {Génot, V.}, {Georgoulis, M. K.}, {Gissot, S.}, {Giunta, A.}, {Gizon, L.}, {Gómez-Herrero, R.}, {Gontikakis, C.}, {Graham, G.}, {Green, L.}, {Grundy, T.}, {Haberreiter, M.}, {Harra, L. K.}, {Hassler, D. M.}, {Hirzberger, J.}, {Ho, G. C.}, {Hurford, G.}, {Innes, D.}, {Issautier, K.}, {James, A. W.}, {Janitzek, N.}, {Janvier, M.}, {Jeffrey, N.}, {Jenkins, J.}, {Khotyaintsev, Y.}, {Klein, K.-L.}, {Kontar, E. P.}, {Kontogiannis, I.}, {Krafft, C.}, {Krasnoselskikh, V.}, {Kretzschmar, M.}, {Labrosse, N.}, {Lagg, A.}, {Landini, F.}, {Lavraud, B.}, {Leon, I.}, {Lepri, S. T.}, {Lewis, G. R.}, {Liewer, P.}, {Linker, J.}, {Livi, S.}, {Long, D. M.}, {Louarn, P.}, {Malandraki, O.}, {Maloney, S.}, {Martinez-Pillet, V.}, {Martinovic, M.}, {Masson, A.}, {Matthews, S.}, {Matteini, L.}, {Meyer-Vernet, N.}, {Moraitis, K.},
  {Morton, R. J.}, {Musset, S.}, {Nicolaou, G.}, {Nindos, A.}, {O’Brien, H.}, {Orozco Suarez, D.}, {Owens, M.}, {Pancrazzi, M.}, {Papaioannou, A.}, {Parenti, S.}, {Pariat, E.}, {Patsourakos, S.}, {Perrone, D.}, {Peter, H.}, {Pinto, R. F.}, {Plainaki, C.}, {Plettemeier, D.}, {Plunkett, S. P.}, {Raines, J. M.}, {Raouafi, N.}, {Reid, H.}, {Retino, A.}, {Rezeau, L.}, {Rochus, P.}, {Rodriguez, L.}, {Rodriguez-Garcia, L.}, {Roth, M.}, {Rouillard, A. P.}, {Sahraoui, F.}, {Sasso, C.}, {Schou, J.}, {Schühle, U.}, {Sorriso-Valvo, L.}, {Soucek, J.}, {Spadaro, D.}, {Stangalini, M.}, {Stansby, D.}, {Steller, M.}, {Strugarek, A.}, {Štverák, Š.}, {Susino, R.}, {Telloni, D.}, {Terasa, C.}, {Teriaca, L.}, {Toledo-Redondo, S.}, {del Toro Iniesta, J. C.}, {Tsiropoula, G.}, {Tsounis, A.}, {Tziotziou, K.}, {Valentini, F.}, {Vaivads, A.}, {Vecchio, A.}, {Velli, M.}, {Verbeeck, C.}, {Verdini, A.}, {Verscharen, D.}, {Vilmer, N.}, {Vourlidas, A.}, {Wicks, R.}, {Wimmer-Schweingruber, R. F.}, {Wiegelmann, T.}, {Young, P. R.}, \&
  {Zhukov, A. N.}}]{Zouganelis2020}
{Zouganelis, I.}, {De Groof, A.}, {Walsh, A. P.}, {et~al.} 2020, A\&A, 642, A3

\end{thebibliography}

\begin{appendix}
\section{Degradation correction}
\begin{figure}[h!tbp]
   \centering
   \includegraphics[scale=0.095]{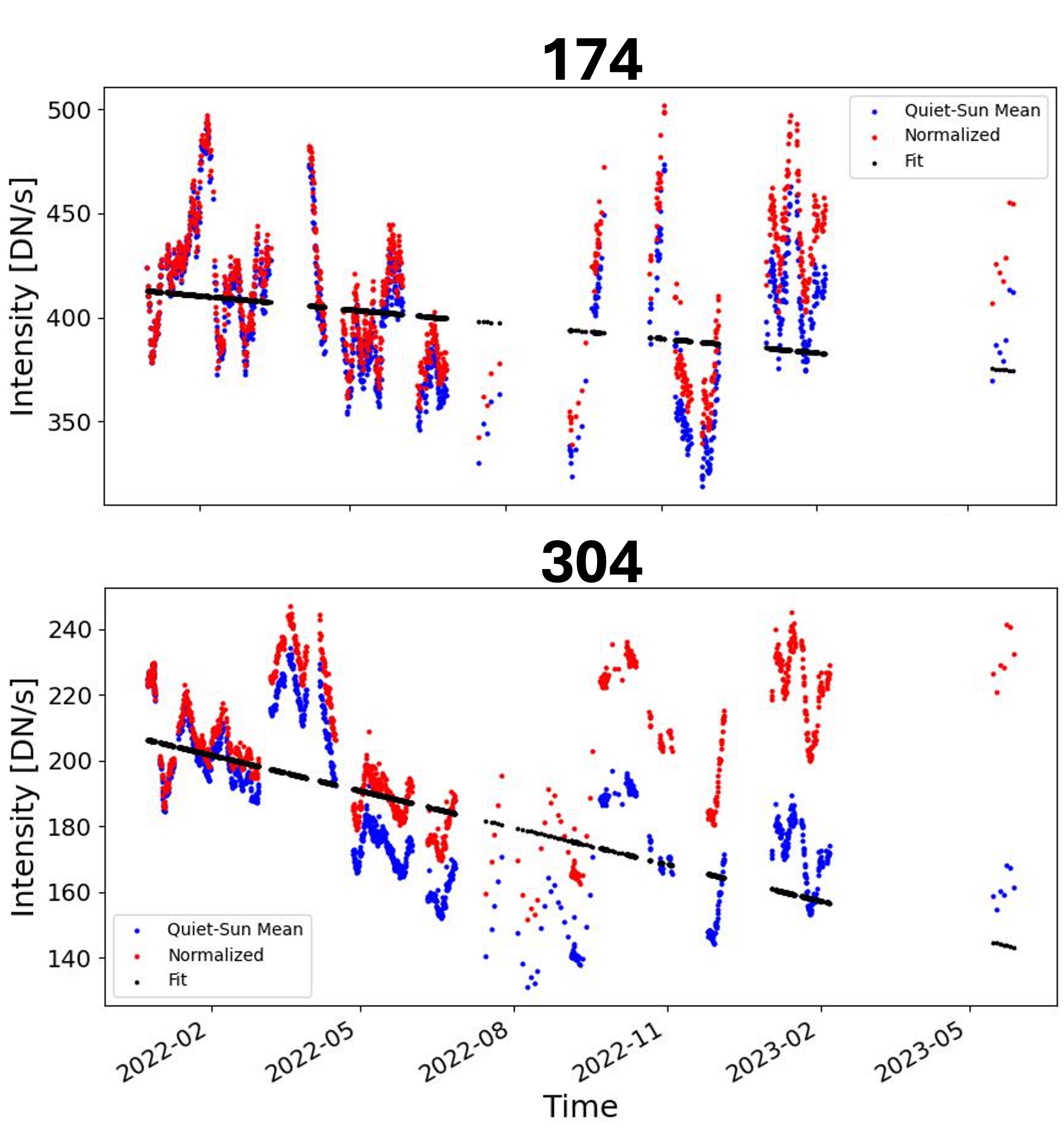}
   \caption{Degradation correction for the EUI/FSI 174~\AA~(first panel) and 304~\AA~channel (second panel) over the training set from January 2022 to May 2023. The blue line corresponds to the quiet Sun mean intensities, the black line to the linear fit and the red line to the corrected light curve.}
\label{fig:deg_corr}
\end{figure}

We perform a degradation correction analysis for the EUI/FSI instrument as described in Sect.\,\ref{sec:fsi_data}. The analysis is performed over the training set from January 2022 to May 2023 and is shown in Fig.\,\ref{fig:deg_corr}. We extract the quiet Sun from full-Sun observations using a thresholding method. The threshold is set according to the median intensity plus the standard deviation of each observation. All regions with values below this threshold are considered to represent the quiet Sun. In Fig.\,\ref{fig:deg_corr} we plot the quiet Sun mean intensities in blue, the linear fit in black and the corrected light curve in red for both the the 174~\AA~and the 304~\AA~channel.  We have performed the degradation analysis up to the latest available data in 2025. However, the degradation fit parameters do not change significantly. Consequently, the degradation correction until 2023 is sufficient to model the degradation up to 2025.

\section{Histogram comparison}
\begin{figure}[h!tbp]
   \centering
   \includegraphics[scale=0.05]{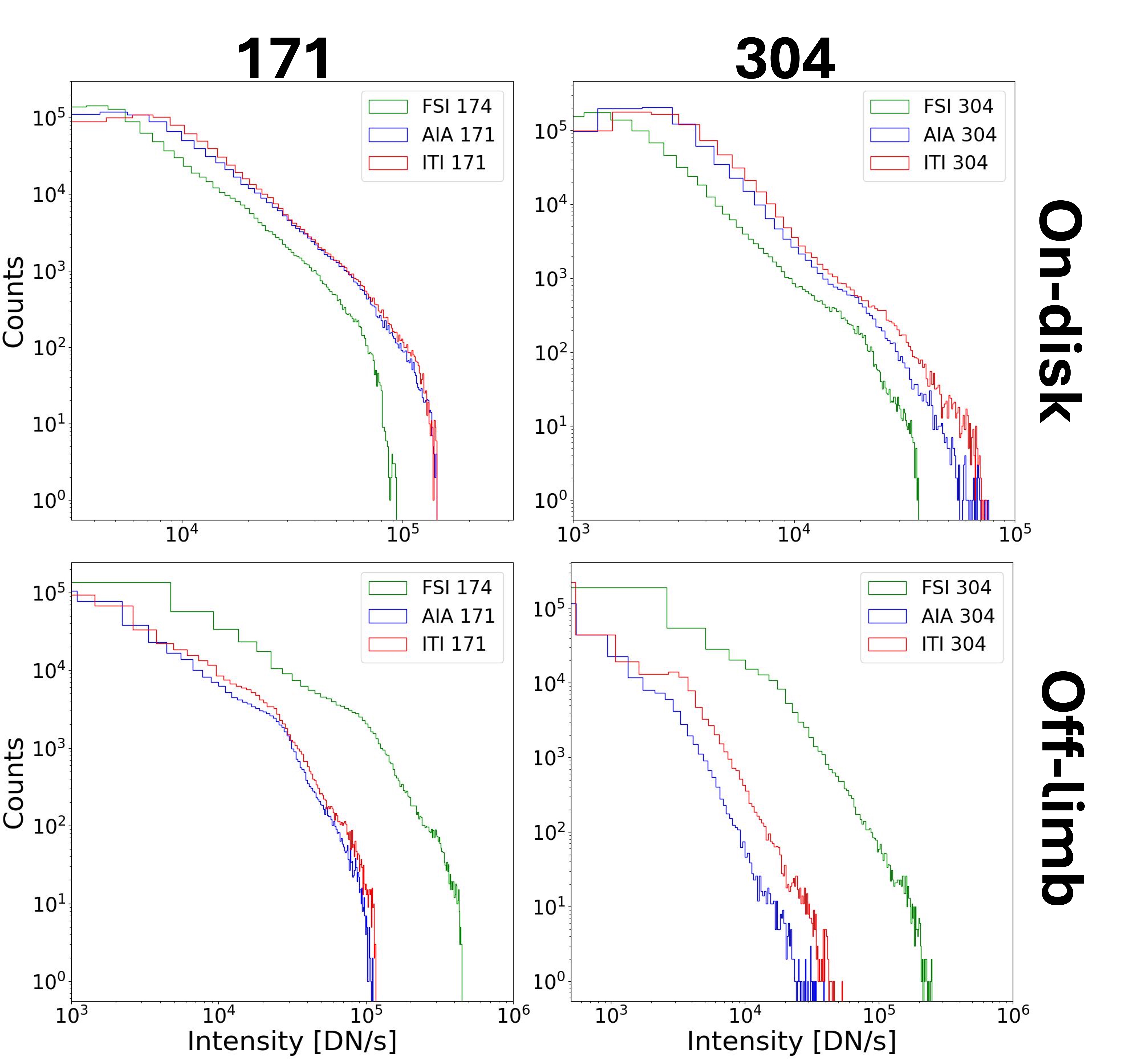}
   \caption{Comparison of the logarithmic intensities for AIA (blue), ITI (red) and the baseline calibrated FSI (green) over test set 1 (Earth aligned). The first row shows the on-disk histogram and the second row the off-limb histogram for the 171~\AA~and the 304~\AA~channel.}
\label{fig:hist_comp_off_disk}
\end{figure}

To show that ITI can provide robust image calibration we calculate the histograms for 171~\AA~and 304~\AA~over test set 1 (Earth aligned) for the on-disk and off-limb pixels. In Fig\,\ref{fig:hist_comp_off_disk} the first row shows the on-disk histograms on a logarithmic scale with AIA in blue, ITI in red and the baseline FSI in green, same as in Fig.\ref{fig:histogram}. The second row shows the off-limb histogram. For both, ITI (red) matches the AIA distribution (blue) closer than the baseline calibrated FSI observations (green).

\end{appendix}
\end{document}